\documentclass[sigconf]{acmart}
\AtBeginDocument{%
  \providecommand\BibTeX{{%
    \normalfont B\kern-0.5em{\scshape i\kern-0.25em b}\kern-0.8em\TeX}}}

\setcopyright{acmcopyright}
\copyrightyear{2018}
\acmYear{2018}
\acmDOI{10.1145/1122445.1122456}

\acmConference[Woodstock '18]{Woodstock '18: ACM Symposium on Neural
  Gaze Detection}{June 03--05, 2018}{Woodstock, NY}
\acmBooktitle{Woodstock '18: ACM Symposium on Neural Gaze Detection,
  June 03--05, 2018, Woodstock, NY}
\acmPrice{15.00}
\acmISBN{978-1-4503-XXXX-X/18/06}

\begin{document}
\title{Corrective Information Does Not Necessarily Curb Social Disruption}

\author{Ryusuke Iizuka}
\affiliation{%
  \institution{The University of Tokyo}
  \country{Japan}
}
\email{iizuka@torilab.net}

\author{Fujio Toriumi}
\affiliation{%
  \institution{The University of Tokyo}
  \country{Japan}
}
\email{tori@sys.t.u-tokyo.ac.jp}

\author{Mao Nishiguchi}
\affiliation{%
  \institution{The University of Tokyo}
  \country{Japan}
}
\email{mao-nishiguchi@g.ecc.u-tokyo.ac.jp}

\author{Masanori Takano}
\affiliation{%
  \institution{CyberAgent, Inc.}
  \country{Japan}
}
\email{takano_masanori@cyberagent.co.jp}

\author{Mitsuo Yoshida}
\affiliation{%
  \institution{Toyohashi University of Technology}
  \country{Japan}
}
\email{yoshida@cs.tut.ac.jp}

\begin{abstract}
The spread of misinformation can cause social confusion.
The authenticity of information on a social networking service (SNS) is unknown, and false information can be easily spread.
Consequently, many studies have been conducted on methods to control the spread of misinformation on SNS. However, few studies have examined the impact of the spread of misinformation and its corrections on society.
This study models the impact of the reduction of misinformation and the diffusion of corrective information on social disruption, and it identifies the features of this impact.
In this study, we analyzed misinformation regarding the shortage of toilet paper during the 2020 COVID-19 epidemic, its corrections, and the excessive purchasing caused by this information.
First, we analyze the amount of misinformation and corrective information spread on SNS, and we create a regression model to estimate the real-world impact of misinformation and its correction. This model is used to analyze the change in real-world impact corresponding to the change in the diffusion of misinformation and corrective information.
Our analysis shows that the corrective information was spread to a much greater extent than the misinformation. In addition, our model reveals that the corrective information was what caused the excessive purchasing behavior.
As a result of our further analysis, we found that the amount of diffusion of corrective information required to minimize the impact on the real world depends on the amount of the diffusion of misinformation.
\end{abstract}




\maketitle

\section{Introduction}
Today, social networking service (SNS) such as Twitter and Facebook are used by a large number of people for various purposes, such as information collection and communication.
SNS is convenient because it's easy to gather a large volume of information, but there remains a problem with the accuracy of the obtained information.
The spread of falsehoods and misinformation can sometimes cause social confusion and serious consequences.

During the COVID-19 pandemic of 2020, a variety of information was spread on SNS, including a large amount of misinformation.
In fact, there have been a number of false claims regarding coronaviruses, such as its supposed vulnerability to heat and the effectiveness of drinking lots of hot water, at 26-27 degrees Celsius, in preventing infection. Another false claim was that the WHO had changed its opinion to the position that quarantine of infected people is unnecessary.
During the second half of February and the first half of March 2020, information was spread that claimed or denied that there would be a shortage of toilet paper due to the coronavirus epidemic.
As a result, it was difficult to obtain toilet paper.


Many studies have been conducted on the spread of misinformation on SNS.
Much of the spread of misinformation is known, including that in the 2016 U.S. presidential election, when American adults may have seen one or more fake news stories in the months before the election \cite{fake_news_1}.
Studies on how to reduce the spread of false rumors have been conducted using simulations, which have shown the effectiveness of strategies that encourage users with many followers to spread corrective information\cite{prevent}.
The effect of additional information aimed at correcting erroneous information has been studied, including which fields of information are most likely to be corrected, which people are most likely to accept correction, and what kind of content is most likely to be corrected\cite{effect_1}\cite{effect_2}.
On the other hand, it is also known that, depending on the nature of the target audience, the effort to correct information can actually be counterproductive \cite{effect_3}.
These studies focus on the spread of falsehoods and corrective information on SNS. However, what we should really aim for is to prevent the social disruption caused by falsehood and corrective information.

The purpose of this study is to clarify the relationship between the spread of false rumors and corrective information on SNS and the confusion these linked behaviors cause in the real world.
This study focuses on the information spread regarding a purported toilet paper shortage and the excessive purchases of toilet paper this information caused.
The first step in our analysis was to clarify how many falsehoods and corrections related to the toilet paper shortage were spread through Twitter data. Next, using sales data from supermarkets, we modeled the relationship between information diffusion and purchased volume using regression analysis.
We then analyzed how much of the corrective information should be disseminated in order to reduce the purchase volume.

The contributions of this study are as follows.
\begin{itemize}
    \item We estimated the actual impact of misinformation and its correction on society, and we showed that corrective information does not necessarily reduce social confusion.
    \item We developed a model to estimate the impact of misinformation and their corrections on excessive purchasing behavior in the real world.
    \item We present one possible guidelines for taking the information-dissemination measures necessary to minimize excessive purchasing behavior.
    \item We show how much corrective information should be spread in relation to the degree of diffusion of misinformation.
\end{itemize}

This paper is organized as follows. First, related studies are described in Section 2.
Section 3 gives details on the data used; Section 4 presents our analysis of the impact of misinformation on sales; Section 5 discusses measures to reduce the diffusion of corrective information and its impact; and Section 6 presents an estimation of the optimal diffusion rate of corrective information based on the misinformation diffusion rate.
Finally, Section 7 gives our conclusion.
\
\section{Related Work}

\subsection{The spread of misinformation}
Here, we present a study on the spread of misinformation.
H. Allcott et al.\cite{fake_news_1} conducted an analysis of fake news during the 2016 US presidential election. Their results suggest that U.S. adults may have seen one or more news articles in the months before the election.

Chengcheng Shao et al.\cite{fake_news_3} also analyzed the same events. Their results show that social bots played a significant role in the spread of hoaxes and that the accounts that actively spread hoaxes had a high probability of being bots.

Karishma Sharma et al. \cite{fake_news_4} provided an analysis of the spread of COVID-19 misinformation and showed that misinformation on COVID-19 was spread across national borders on SNS.

Soroush Vosoughi et al.\cite{fake_news_2} analyzed the spread of correct, partially incorrect, and misinformation on Twitter. As a result, they found that misinformation was spread more quickly and widely than the other types of information. They also found that political misinformation was more widely disseminated than other categories of misinformation.

As mentioned above, it has been shown that the spread of falsehoods occurs in various fields, and the existence of bots is sometimes behind the spread of falsehoods. It has also been shown that misinformation is more likely to spread than other kinds of information.

\subsection{Preventing the spread of misinformation}
Here, we introduce previous research on preventing the spread of misinformation, including that by members of our group. Y. Okada et al.\cite{prevent} developed a model for the diffusion of misinformation based on the SIR model and devised a strategy to stop the spread of misinformation. As a result, they found that the strategy of spreading corrective information to users with many followers was effective.

J. Kim et al.\cite{prevent_2} proposed a crowd-based online algorithm to detect and prevent the spread of misinformation through SNS. They used data from Twitter and Webio to show that the proposed method could be effective.

In addition, to prevent the spread of misinformation, it is necessary to first detect it.
Various features are used to detect fake news.
According to Kai shu et al.\cite{summary} the features used to detect fake news are either based on the news content or based on the social context.
Features based on news content are divided into linguistic-based features and visual-based features.
Linguistic-based features include words and sentences and so on. Aditi Gupta et al.\cite{prevent_4} attempted to use these features to distinguish hyperpartisan news and fake news.
Other studies using these features include those by Sadia Afroz et al.\cite{ling_additional}.

Visual-based features include photos and videos.
Martin Potthast et al.\cite{prevent_3} classified assumed images of Hurricane Sandy into false images and actual images, and they showed a classification accuracy of 97\%.
Other studies using these features include those by Zhiwei Jin et al.\cite{visual_additional}.

There have been many studies on social context-based features such as user-based features (age and number of tweets, etc.)\cite{additional_1}, post-based features(emotions or opinions towards fake news)\cite{additional_2} and network-based features\cite{additional_3}.


Since bots may be involved in the spread of falsehoods as described above, many studies on the detection of bots have been carried out.
For bot detection, a method using supervised learning is known to produce good results \cite{supervised_1}\cite{supervised_2}.
As a method using unsupervised learning, Nikan Chavoshi et al. \cite{unsupervised_1} applied dynamic time warping to detect bots based on the correlation between user accounts, and the results were better than those obtained by supervised learning.

Many research works have attempted to prevent the spread of misinformation as described above, such as opposing it with corrective information and developing ways to detect misinformation.
However, these approaches have not taken into account the issue of whether the corrective information would really stop the spread of misinformation.

\subsection{The effect of correcting misinformation}
Here, we present a study on the effect of corrective information.
B. Nyhan et al.\cite{effect_3} conducted a survey of parents with children aged 17 years or younger to determine whether corrective information would encourage vaccination.
As a result, none of the corrective information used in the survey encouraged vaccination. 
Subjects who were most reluctant to be vaccinated were shown to become even more reluctant toward vaccination.

J. M. Carey et al.\cite{effect_4} analyzed whether corrective information on the Zika and yellow fever hoaxes in Brazil would be effective in dispelling misconceptions. Corrective information about Zika fever failed to correct the misconceptions about Zika fever, but it was found to be effective for the better known yellow fever.

Accordingly, the effectiveness of corrective information has been studied in various ways, but it depends on various factors such as the type of target hoax. Therefore, the following meta-analysis was conducted.
M.-p. S. Chan et al.\cite{effect_1} analyzed the conditions under which corrective information is effective in clearing up the misconceptions of those who believe false information. As a result, they found, among other things, that it was effective for those who were more likely to refute a hoax but less effective for those who had created reasons to believe the misinformation.

N. Walter et al. \cite{effect_2} conducted a meta-analysis of the effect of corrective information on correcting erroneous perceptions and the factors that influence the effect. The effectiveness of the corrective information was moderate, and they found that it was more difficult to correct misinformation about politics and marketing than that about health. Furthermore, real-world misinformation, such as climate change denial, is more difficult to correct than constructed misinformation, such as plane crashes.

A lot of research has been done on whether the correction of misinformation can solve the problem of misrecognition. It is known that the effectiveness of misinformation varies depending on such factors as the type of misinformation and the nature of the person who believes the misinformation.
However, we look not at the effect of unraveling misperceptions, but at their impact on social disruption.
\

\section{Data}
\subsection{Japan's toilet paper shortage misinformation causes social confusion}
Between late February and early March 2020, false claims of toilet paper shortages and information that denied the claims were spread on SNS. As a result, although there was plenty of inventory at the factory, logistics could not keep up with demand, and toilet paper disappeared temporarily from stores in many places. This situation led Prime Minister Abe to deny the misinformation and call for calm purchasing behavior.

\subsection{Tweets Data}
We used two major types of data from Twitter in this study: tweet data and follow-relationship data.
To get tweet data, we targeted tweets related to the shortage of toilet paper.
The data were retrieved based on the related keywords in Japanese, which were "toilet paper," "tissue," "sanitary napkins," and "Act on Emergency Measures for Stabilizing Living Conditions of the Public."
We classified the data into three categories according to the content of the tweets. Classification was done manually. The three categories are as follows.
\begin{itemize}
    \item Content that encourages people to buy toilet paper or hoaxes about lack of toilet paper\ (misinformation tweets)
    \item Content that indicates denial of hoaxes or denial of lack of toilet paper (corrective tweets)
    \item Content indicating that toilet paper is actually sold out (sold-out tweets)
\end{itemize}
The data were collected from February 21 to March 10. The total number of tweets obtained during the collection period was 4,476,754, of which the total number of RTs was 2,945,782.
A summary of the data for each type of content is shown in Table \ref{tbl:data}.
The reason why we included only misinformation tweets in the top 6,000 RTs is because we wanted to collect as many misinformation tweets as possible, since the number of misinformation tweets that were widely spread was relatively small.

\begin{table*}[hbt]
  \caption{Data Overview}
  \label{tbl:data}
  \begin{tabular}{|c|c|c|c|} \hline
    Contents & requirement & 
    \begin{tabular}{c}
     Number of tweets \\(no duplication)  
    \end{tabular} 
    & Number of users who retweet \\ \hline
     misinformation tweet & number of RTs top 6000 tweets & 8 & 712 \\ \hline
     corrective tweet & number of RTs top 1400 tweets & 229 & 332,881 \\ \hline
     sold-out tweet & number of RTs top 1400 tweets & 42 & 18,264 \\ \hline
  \end{tabular}
\end{table*}

In order to estimate the number of people who might have seen the misinformation tweets and corrective tweets, we collected the followers of the accounts that retweeted those tweets.
Followers were obtained using the Twitter API to obtain the followers at the time of acquisition.
Note that no information such as tweets of private accounts is available.
As a result, we collected 97,430,525 accounts.

\subsection{Toilet paper sales index}
\begin{figure}[htbp]
\begin{center}
\includegraphics[width=80mm]{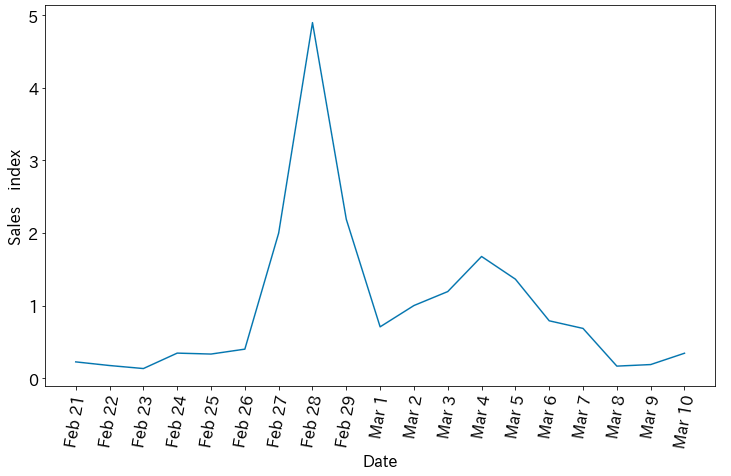}
\caption{Changes in Sales Index}
\label{fig:sales}
\end{center}
\end{figure}

The toilet paper sales index was used as an indicator of how much toilet paper was sold.This data was provided by NOWCAST, INC.
The sales index of toilet paper is an index that expresses the degree of change in sales of toilet paper compared to the previous year, and it is expressed in equation \ref{eq:sales_index}.
S is the sales index and $Sales_t$ is sales of toilet paper on a given day t, dt=364 days.
This index is created from point-of-sale (POS) data from 1200 supermarkets across the country.

\begin{equation}
    \label{eq:sales_index}
    S=(Sales_t/Sales_{t-dt})-1
\end{equation}
The change in the sales index during the data analysis period is shown in Figure \ref{fig:sales}. Throughout this period, sales of toilet paper were higher than usual, with a peak on February 28.


\section{Analysis of the Impact of Tweets on Sales}
\subsection{Number of possible viewers}
To find out how widely the tweets about the supposed toilet paper shortage spread, we checked the number of user who saw each tweet. The following seven types of possible viewers were checked.
\begin{itemize}
    \item Number of possible viewers who see only corrective tweets ($x_1$)
    \item Number of possible viewers who see only misinformation tweets ($x_2$)
    \item Number of possible viewers who see only sold-out tweets ($x_3$)
    \item Number of possible viewers who see corrective tweets and misinformation tweets ($x_4$)
    \item Number of possible viewers who see corrective tweets and sold-out tweets ($x_5$)
    \item Number of possible viewers who see misinformation tweets and sold-out tweets ($x_6$)
    \item Number of possible viewers who see all types of tweets ($x_7$)
\end{itemize}

\begin{figure}[htbp]
\begin{center}
\includegraphics[width=80mm]{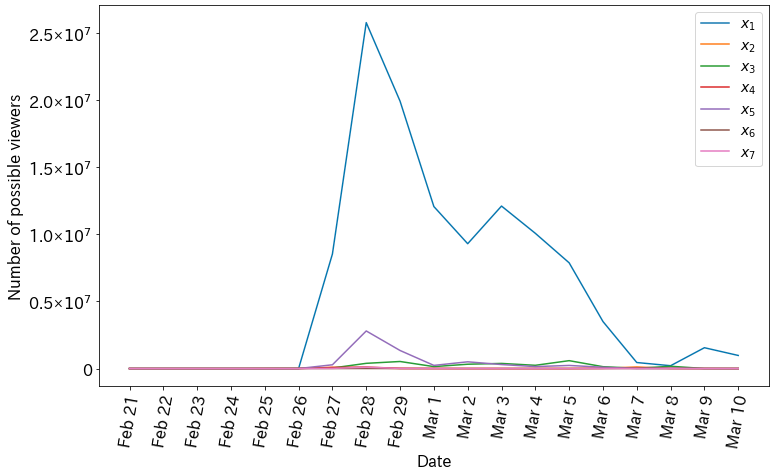}
\caption{Number of possible viewers}
\label{fig:number_1}
\end{center}
\end{figure}

Figure \ref{fig:number_1} and Table \ref{tbl:number_2} show the results of studying the numbers of possible viewers of the above seven types. The number of users who saw only the corrective tweets was the largest, followed by those who saw the corrective tweets and sold-out tweets, and then those who saw only the sold-out tweets. The number of users who saw only misinformation tweets was a mere 0.28\% of the number who saw corrective tweets only.


\begin{table*}[htb]
  \caption{Total estimated number of viewers} 
  \label{tbl:number_2}
  \begin{tabular}{|c|r|} \hline 
    \ &Total estimated number of possible viewers \\ \hline
     only corrective tweets & 112,440,832 \\ \hline
     only misinformation tweets & 311,345 \\ \hline
     only sold-out tweets & 2,974,369 \\ \hline
     corrective tweets and misinformation tweets & 251,100\\ \hline
     corrective tweets and sold-out tweets & 5,967,030\\ \hline
     misinformation tweets and sold-out tweets & 4,157\\ \hline
     all types of tweets & 124,953 \\ \hline
  \end{tabular}
\end{table*}

\subsection{Analysis of the impact of information viewers on sales}
\subsubsection{summary of methodology}
First, we model the daily sales index trends using multiple regression analysis. For the explanatory variables, we use the principal component, which is the number of possible viewers of each type described above as input.

The reason for conducting the principal components analysis is to avoid multicollinearity. Table 3 presents the correlation coefficients between the numbers of people seen for each type of content.

\begin{table}[htb]
  \caption{Correlation coefficients between numbers of users}
  \label{tbl:correlation}
  \begin{tabular}{|c|c|c|c|c|c|c|c|} \hline
    \ &$x_1$ & $x_2$ & $x_3$ & $x_4$ & $x_5$ & $x_6$ & $x_7$ \\ \hline
     $x_1$ & 1.0 & 0.091 & 0.76 & 0.58 & 0.87 & 0.58 & 0.68 \\ \hline
     $x_2$ & 0.091 & 1.0 & -0.15 & 0.68 & 0.18 & 0.45 & 0.29 \\ \hline
     $x_3$ & 0.76 & -0.15 & 1.0 & 0.16 & 0.57 & 0.23 & 0.31 \\ \hline
     $x_4$ & 0.58 & 0.68 & 0.16 & 1.0 & 0.74 & 0.86 & 0.84 \\ \hline
     $x_5$ & 0.87 & 0.18 & 0.57 & 0.74 & 1.0 & 0.83 & 0.91 \\ \hline
     $x_6$ & 0.58 & 0.45 & 0.23 & 0.86 & 0.83 & 1.0 & 0.97 \\ \hline
     $x_7$ & 0.68 & 0.29 & 0.31 & 0.84 & 0.91 & 0.97 & 1.0 \\ \hline
  \end{tabular}
\end{table}

\subsubsection{principal component analysis}
Table \ref{tbl:pca_1} shows the eigenvectors of each principal component. In this table, PC stands for principal components. The eigenvector of the first principal component has a large $x_1$ component, and the eigenvector of the second principal component has a large $x_5$ component. The eigenvector of the third principal component has a large $x_3$ component, and the eigenvector of the fourth principal component has a large $x_2$ component.

\begin{table}[htb]
  \caption{Eigenvectors of principal components}
  \label{tbl:pca_1}
  \begin{tabular}{|c|c|c|c|c|c|c|c|} \hline 
    \ &$x_1$ & $x_2$ & $x_3$ & $x_4$ & $x_5$ & $x_6$ & $x_7$ \\ \hline
     1st PC & 0.99 & 0.00 & 0.02 & 0.00 & 0.08 & 0.00 & 0.00 \\ \hline
     2nd PC & -0.07 & 0.02 & -0.13 & 0.04 & 0.99 & 0.00 & 0.04 \\ \hline
     3rd PC & -0.03 & -0.08 & 0.98 & -0.08 & 0.13 & 0.00 & -0.02 \\ \hline
     4th PC & 0.00 & 0.89 & 0.11 & 0.43 & -0.02 & 0.00 & 0.05 \\ \hline
     5th PC & 0.00 & 0.41 & -0.04 & -0.76 & 0.04 & -0.01 & -0.51 \\ \hline
     6th PC & 0.00 & -0.18 & 0.00 & 0.48 & 0.02 & -0.03 & -0.86 \\ \hline
     7th PC & 0.00 & 0.00 & 0.00 & 0.00 & 0.00 & 0.99 & -0.03 \\ \hline
  \end{tabular}
\end{table}

Table \ref{tbl:pca_2} shows the contribution ratio of each principal component.
The contribution rate is represented by equation \ref{eq:contribution}, where C is the contribution rate, p is the number of principal components, and ${l}_i$ is the eigenvalue of the i-th principal component.
\begin{equation}
    \label{eq:contribution}
    C=\frac{l_i}{\sum_{i=1}^{p} l_i}
\end{equation}
The contribution ratio indicates the percentage of the amount of information accounted for by the corresponding principal component out of the total amount of information. The contribution of the first principal component is very large at 9.97$\times$ $10^{-1}$, indicating that the first principal component can explain most of the original data. In addition, the eigenvector of the first principal component has a large $x_1$ component, indicating that $x_1$ can explain most of the original data. 

\begin{table}[htb]
  \caption{Contribution of each principal component}
  \label{tbl:pca_2}
  \begin{tabular}{|c|c|} \hline 
    \ &Contribution ratio \\ \hline
     1st PC & 9.97 $\times$ $10^{-1}$ \\ \hline
     2nd PC & 1.83 $\times$ $10^{-3}$ \\ \hline
     3rd PC & 2.35 $\times$ $10^{-4}$ \\ \hline
     4th PC & 2.03 $\times$ $10^{-5}$ \\ \hline
     5th PC & 1.72 $\times$ $10^{-6}$ \\ \hline
     6th PC & 3.43 $\times$ $10^{-7}$ \\ \hline
     7th PC & 8.44 $\times$ $10^{-11}$ \\ \hline
  \end{tabular}
\end{table}

\subsubsection{multiple regression analysis} 
Table \ref{tbl:lr} gives the results of multiple regression analysis, showing the coefficients and p-values.
$a_1$, $a_2$, $a_3$ and $a_4$ are regression coefficients, and $b$ is the constant term.
Only the third principal component has a p-value greater than 0.05.
The coefficients and t-values are presented in Table \ref{tbl:importance_ind} and Table\ref{tbl:importance_group} respectively, because the regression analysis is done after the principal component analysis.
 


\begin{table*}[htb]
  \caption{Regression coefficients and p-values}
  \label{tbl:lr}
  \begin{tabular}{|c|c|c|c|c|c|} \hline 
    \ &$a_1$ & $a_2$ & $a_3$ & $a_4$ & b \\ \hline
     coefficients & 1.35$\times$ $10^{-7}$ & 10.13$\times$ $10^{-7}$& -2.494$\times$ $10^{-7}$ & 69.53$\times$ $10^{-7}$ & 0.9919 \\ \hline
     p-values & 0.000 & 0.001 & 0.695 & 0.006 & 0.000 \\ \hline
     t-values & 13.55 & 4.470 & -0.400 & 3.211 & 13.47 \\ \hline
  \end{tabular}
\end{table*}

Table \ref{tbl:importance_ind} shows the inner product of the regression coefficients and eigenvectors.
The inner product of the regression coefficients and eigenvectors indicates the importance of each variable and, in this case, the magnitude of its impact on the sales index per user.
We can see that the impact per user is greatest for the number of possible viewers who only saw the misinformation($x_2$).
In other words, if the number of users who saw only the misinformation increased, the purchase of toilet paper would accelerate and there would likely be a shortage.
We also found that the next most influential accounts were $x_4$, i.e., those who saw both misinformation and corrective information.
This is because these accounts are the only accounts that may have seen information on both sides, and thus they may have actually included many accounts that saw only misinformation.


\begin{table}[htb]
  \caption{Impact on sales index per possible viewer}
  \label{tbl:importance_ind}
  \begin{tabular}{|c|r|} \hline 
    \ & Impact on sales index \\ \hline
     $x_1$ & 5.35 $\times$ $10^{-8}$ \\ \hline
     $x_2$ & 624.00 $\times$ $10^{-8}$ \\ \hline
     $x_3$ & 44.80 $\times$ $10^{-8}$ \\ \hline
     $x_4$ & 309.00 $\times$ $10^{-8}$ \\ \hline
     $x_5$ & 79.20 $\times$ $10^{-8}$ \\ \hline
     $x_6$ & 2.88 $\times$ $10^{-8}$ \\ \hline
     $x_7$ & 43.10 $\times$ $10^{-8}$ \\ \hline
  \end{tabular}
\end{table}

\begin{figure}[htbp]
\begin{center}
\includegraphics[width=80mm]{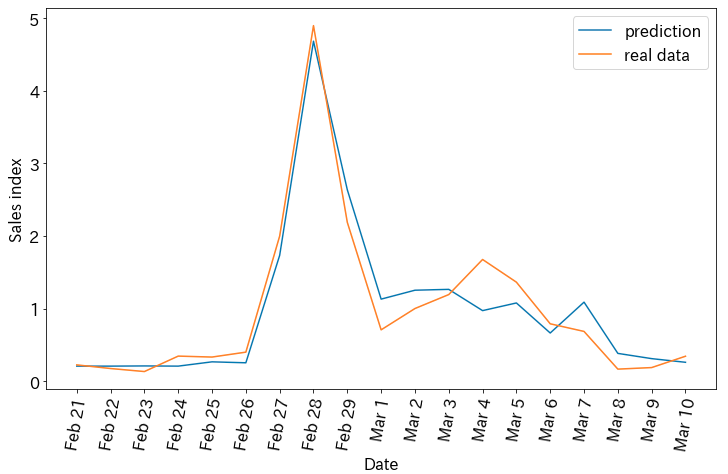}
\caption{Regressed sales index}
\label{fig:lr_1}
\end{center}
\end{figure}

Figure \ref{fig:lr_1} shows the evolution of the regressed and actual sales indexes. Although there are small deviations, the regression is accurate, with a coefficient of determination of 0.939 and F value of 53.49.
\

Table\ref{tbl:importance_group} shows the results obtained by multiplying the total estimated number of possible viewers shown in Table\ref{tbl:number_2} by the impact on sales index per possible viewer.
$s_{x_i}$ is the results obtained for each of $x_1$ to $x_7$
The results show which types of accounts affected sales and by how much, showing that the type of accounts that saw only corrective tweets were assumed to have the most impact on sales.
This result is also consistent with Table\ref{tbl:pca_2} and Table\ref{tbl:lr}.


\begin{table}[htb]
  \caption{Impact on sales index per type of accounts}
  \label{tbl:importance_group}
  \begin{tabular}{|c|r|} \hline 
    \ & Impact on sales index \\ \hline
     $s_{x_1}$ & 6.0236 \\ \hline
     $s_{x_2}$ & 1.9412 \\ \hline
     $s_{x_3}$ & 1.3339\\ \hline
     $s_{x_4}$ & 0.7763 \\ \hline
     $s_{x_5}$ & 4.727 \\ \hline
     $s_{x_6}$ & 0.0001 \\ \hline
     $s_{x_7}$ & 0.0538 \\ \hline
  \end{tabular}
\end{table}

\section{Impact of reducing the spread of corrective information}
\subsection{Method of reducing the spread of corrective tweets}

Misinformation and its corrections do not necessarily reduce social confusion if the corrections are widespread.
In this section, we discuss how to reduce the social confusion in Japan by using a model that estimates the impact of the toilet paper shortage as discussed in Section 4.


First, we simulate the reduction of corrective tweets based on the real data.
Here, we assume that a percentage of users who retweeted corrective tweets in the real data do not RT.
In addition, we assume that the users who lost the opportunity to see the retweeted tweets do not retweet, even though they had retweeted the corrective tweets in the real data.

\subsection{Experiment on reducing the spread of corrective tweets}
\subsubsection{experimental conditions}
Here, we use actual data to examine the change in the sales index when the number of users who retweeted corrective tweets is reduced.
In this experiment, we use the parameter of what percentage of users who saw the corrective tweets retweeted the corrective tweets (hereinafter referred to as the "corrective RT rate").
The corrective RT rate was set at seven levels: 0.79\%,0.63\%,0.47\%,0.32\%,0.16\% and 0.0\%,
and the numbers of users with these corrective RT rates represent 100\%, 80\%, 60\%, 40\%, 20\%, and 0\% of the users who retweeted the misinformation in the real data, respectively.
\subsubsection{sales forecast}
\begin{figure}[htbp]
\begin{center}
\includegraphics[width=80mm]{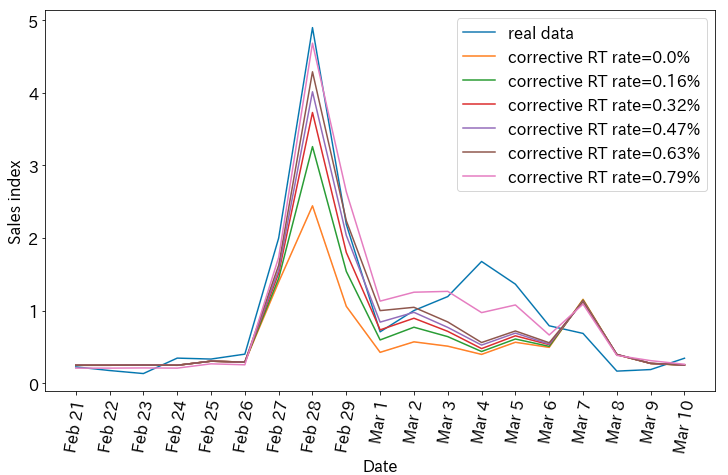}
\caption{Sales index estimated in experiment on reducing the spread of corrective tweets}
\label{fig:shrink}
\end{center}
\end{figure}

Figure \ref{fig:shrink} shows the results of the experiment on reducing the spread of corrective tweets. The smaller the corrective RT rate, the smaller the sales index is estimated to be.
We found that the sales index is estimated to be the smallest when no one retweets the corrective tweet.\par
Realistically, however, we expect the spread of misinformation tweets to increase when there are fewer corrective tweets. In Section 6, we present an experiment that takes this expectation into account.


\subsection{Preventing the spread of information by misinformation in non-contact users}
In the experiment on reducing the spread of corrective tweets, we randomly selected users to retweet the corrective tweets.
However, it is not realistic to arbitrarily increase or decrease the number of users randomly retweeting a corrective tweet.
Therefore, we propose the following guideline for retweeting.
`Only users who saw the misinformation before retweeting the corrective tweet will retweet the corrective tweet.'
Based on this guideline, we estimate the sales index.
\subsubsection{experimental conditions}
In this experiment, there is a parameter that indicates the percentage of users who only see misinformation tweets and who will retweet those misinformation tweets (hereinafter referred to as the “misinformation RT rate").
Since the percentage of users who RT misinformation tweets was 0.186\% among those who saw only misinformation tweets in the actual data, we set the misinformation RT rate as 0.186\% in this experiment.
The corrective RT rates were set to 0.79\%, 0.63\%, 0.47\%, 0.32\%, 0.16\% and 0.0\%, as above.
The number of users who retweeted the corrective tweets is 0.029\% of those who originally retweeted the corrective tweets.
Ten experiments were conducted for seven different conditions, including the six different corrective RT rates and the proposed guideline.

\subsubsection{sales forecast}
Table \ref{tbl:result_3} shows the results of preventing the spread of information by misinformation in non-contact users.
Sum of sales index means total sales index for the period under review.
The proposed guideline shows that sales index are kept small compared to the corrective RT rate of > = 0.16\%,
since few users retweet the corrective tweets, which is consistent with the discussion in the previous paragraph.
Compared with the original sales index (18.85), the proposed method is able to reduce the sales index by 40.4\%.

\begin{table}[htbp]
  \caption{Results of preventing the spread of information by misinformation in non-contact users (average of 10 trials with different random seeds)}
  \label{tbl:result_3}
  \begin{tabular}{|c|c|} \hline 
     Corrective RT rate & Sum of sales index \\ \hline
     $0.029\%$(proposed method) & 11.23 \\ \hline
     $0.0\%$ & 9.192 \\ \hline
     $0.16\%$ & 12.50 \\ \hline
     $0.32\%$ & 13.34 \\ \hline
     $0.47\%$ & 14.11 \\ \hline
     $0.63\%$ & 14.69 \\ \hline
     $0.79\%$ & 15.93 \\ \hline
  \end{tabular}
\end{table}

\section{Estimation of Optimal Correction Diffusion Rate Based on Misinformation Diffusion Rate}
\subsection{Impact of the spread of corrective information on the amount of misinformation spread and sales}
\begin{figure}[htbp]
\begin{center}
\includegraphics[width=80mm]{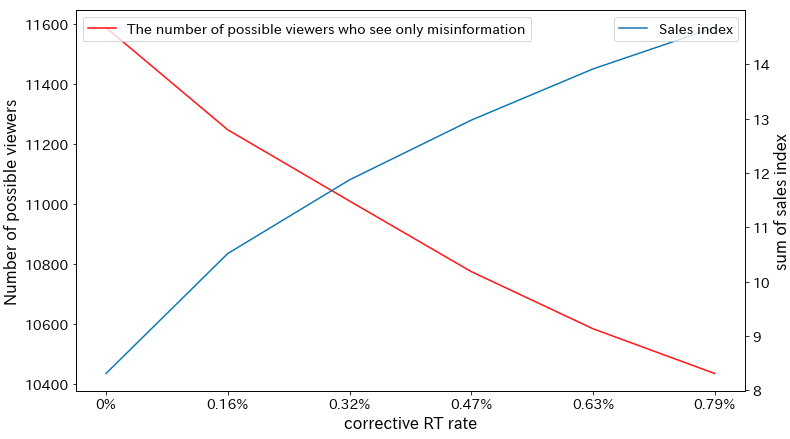}
\caption{Impact of corrective tweets (average of 10 trials with different random seeds)}
\label{fig:result_6_1}
\end{center}
\end{figure}

Figure \ref{fig:result_6_1} shows the relationship between the corrective RT rate, the number of possible viewers who saw the corrective tweets only, and sum of the sales index when the misinformation RT rate is 0.
The vertical line shows the total sales index for the period under review.
We found that as the corrective RT rate increases, the number of possible viewers who saw only the misinformation decreases but the sales index increases.
These results show that, depending on the degree of diffusion, spreading a large amount of corrective information is counterproductive for reducing social confusion.
Therefore, in the following, we investigate the appropriate diffusion of the corrective tweets to the degree of diffusion of the misinformation tweets.


\subsection{Experimental conditions}
The diffusion rate of misinformation tweets and corrective tweets is likely to vary based on the level of credibility.
If a piece of misinformation is too ridiculous or unimportant, it is unlikely to be spread, while misinformation that is highly credible or appears to be important will more likely be spread.
The same is true for the corrective information.
If we assume that the appropriate diffusion rate of corrective information varies with the diffusion power of misinformation, then how corrective information should be diffused will change based on the diffusion rate of the original misinformation.
Therefore, in this section, we clarify the relationship between the diffusion rate of misinformation and the optimal diffusion rate of corrective information.

In this experiment, the corrective RT rates were set to 0.79\%, 0.63\%, 0.47\%, 0.32\%, 0.16\% and 0.0\%.
The misinformation RT rates were then set to 0\%, 0.186\%, 1.0\%, 2.0\%, 3.0\%, 4.0\% and 5.0\%.
A total of 42 combinations of these conditions were tested. 
Ten experiments were conducted for each experimental condition, and the results are the average of the sum of the estimated sales index during the period of interest.

\subsection{Analysis of the impact of corrective tweets on sales}
\begin{figure}[htbp]
\begin{center}
\includegraphics[width=80mm]{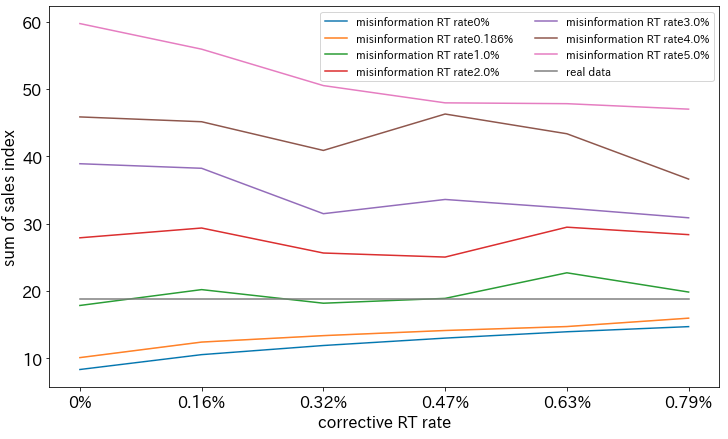}
\caption{Results of analysis of the impact of corrective tweets on sales (average of 10 trials with different random seeds)}
\label{fig:result_1_3}
\end{center}
\end{figure}

\ \\
Figure \ref{fig:result_1_3} shows the results of the analysis of the impact of corrective tweets on sales. The vertical line shows the total sales index for the period under review.

When the misinformation RT rate is small (misinformation RT rate = 0.0\%, 0.186\%), the expected sales increase as the corrective RT rate increases. 
We found that when the misinformation RT rate was large (misinformation RT rate > = 3.0\%), sales decreased as the corrective RT rate increased.

When the spread of misinformation tweets is small, the spread of corrective tweets far outweighs the spread of misinformation, resulting in a larger impact on sales. 
On the other hand, if the spread of misinformation is large, corrective tweets prevent the spread of misinformation tweets and reduce sales.

In terms of actual events, the spread of corrective tweets was counterproductive due to the low misinformation RT rate. However, for more credible and widely spread disinformation, it is useful to spread corrective tweets.

Specifically, in the case used in this study, the spread of corrective tweets was effective when more than $2\%$ of the users who saw only the misinformation tweets retweeted.

Accordingly, as shown in Figure \ref{fig:result_6_1}, an increase in the number of corrective RTs reduced the spread of misinformation, but as shown in Figure \ref{fig:result_1_3}, the corrective information did not necessarily lead directly to a decrease in sales, i.e., social disruption.
That is, the appropriate amount of corrective tweets varies with the misinformation RT rate, i.e., the credibility of the misinformation.

It is usually understood that for misinformation, it is important to correct the misinformation in question as much as possible.
However, it has been shown that excessive corrective information can cause social confusion.
When the misinformation is spread, it is necessary to estimate the impact of the misinformation and the corrective information on society and to carefully consider how to deal with it.

\ 

\section{Conclusion}
In this study, we analyzed the impact of misinformation about a purported toilet paper shortage in Japan during the COVID-19 pandemic of 2020. This analysis was conducted using data on both tweets and sales.

The number of users who were exposed to misinformation and corrective information was calculated by classifying tweets by content.
As a result, the number of possible viewers who saw only the corrective tweets was overwhelmingly high, about 357 times that of possible viewers who saw only the misinformation tweets.

Multiple regression analysis was performed using the number of possible viewers for each tweet explanatory variables.
As a result, the coefficient of determination was 0.939, and F value was 53.49, indicating that sales can be predicted accurately.
The coefficients of the regression equation and other results show that the accounts that had the largest impact on the sales index were the type of accounts that saw only corrective tweets.

\par
We used the model to estimate the sales index when the number of users who retweet corrective tweets changed.
The results show that the sales index decreases as the number of users retweeting a corrective tweet decreases, indicating that corrective tweets are a factor in the excessive purchase of toilet paper.
In this study, we examined the effect on sales of a guideline of "users who did not see the misinformation tweets do not RT the corrective tweets."
The results show that sum of the sales index can be reduced by 40.4\% compared to the actual case, suggesting that this guideline is effective.

\par
Finally, we used the proposed model to estimate the optimal diffusion rate of corrective information (corrective RT rate) when the credibility of misinformation, namely the misinformation diffusion rate  (misinformation RT rate), changes.
The results show that when the credibility of the misinformation is low, i.e., the spread of the misinformation is small, corrective tweets actually promote over-purchasing.
On the other hand, when the credibility of the misinformation is high, i.e., the misinformation rate is high, we show that corrective tweets can reduce over-purchasing.
\ 
\par
While most of the previous studies examined the effect of corrective information on the diffusion rate of misinformation and the contact rate of misinformation, this study examined the effect of corrective information on the reduction of social confusion.

In the future, it will be necessary to identify appropriate corrective strategies based on the rate of spread of misinformation.
In addition, overbuying is not the only disruption that misinformation brings to society. It is also necessary to analyze the factors affecting the social impact of the various types of misinformation as well as to identify the general trends in efforts to optimize the diffusion of corrective information.
Furthermore, it is said that misinformation is not only influenced by social media but also by mass media (Benkler, Yochai, et al., "Mail-In Voter Fraud: Anatomy of a Disinformation Campaign," available at SSRN (2020)).
The establishment of countermeasures against misinformation that take into account influences other than social media remains an important future issue.

\bibliographystyle{ACM-Reference-Format}
\bibliography{acmart}

\end{document}